\documentclass[a4paper,10pt,twoside]{cpc-hepnp}
\usepackage{multicol}
\usepackage{graphicx}
\usepackage{booktabs}
\usepackage{bm}
\usepackage{mathrsfs}
\usepackage{bbm}
\usepackage{amscd}
\usepackage[tbtags]{amsmath}
\usepackage{amssymb,bm,mathrsfs,bbm,amscd}
\usepackage{lastpage}
\begin{document}

\newcommand{\calL}{\ensuremath{\mathcal{L}}}

\fancyhead[co]{\footnotesize BESIII Col.:
{\boldmath Determination of the number of $J/\psi$ events using
inclusive $J/\psi$ decays}} \footnotetext[0]{Received }
\title{\boldmath Determination of the number of $J/\psi$ events
with inclusive $J/\psi$ decays\thanks{Supported in part by National Key Basic Research Program of China under
Contract No. 2015CB856700; National Natural Science Foundation of China
(NSFC) under Contracts Nos. 10805053, 11125525, 11175188, 11235011, 11322544, 11335008,
11425524; the Chinese Academy of Sciences (CAS) Large-Scale Scientific
Facility Program; the CAS Center for Excellence in Particle Physics (CCEPP);
the Collaborative Innovation Center for Particles and Interactions (CICPI);
Joint Large-Scale Scientific Facility Funds of the NSFC and CAS under
Contracts Nos. 11179007, U1232201, U1232107, U1332201; CAS under Contracts Nos.
KJCX2-YW-N29, KJCX2-YW-N45; 100 Talents Program of CAS; INPAC and Shanghai
Key Laboratory for Particle Physics and Cosmology; German Research
Foundation DFG under Contract No. Collaborative Research Center CRC-1044;
Istituto Nazionale di Fisica Nucleare, Italy; Ministry of Development of
Turkey under Contract No. DPT2006K-120470; Russian Foundation for Basic
Research under Contract No. 14-07-91152; U. S. Department of Energy under
Contracts Nos. DE-FG02-04ER41291, DE-FG02-05ER41374, DE-FG02-94ER40823,
DESC0010118; U.S. National Science Foundation; University of Groningen (RuG)
and the Helmholtzzentrum fuer Schwerionenforschung GmbH (GSI), Darmstadt;
WCU Program of National Research Foundation of Korea under Contract No.
R32-2008-000-10155-0
}
\vspace{-0.39in}
}
\maketitle
\author{
M.~Ablikim$^{1}$, M.~N.~Achasov$^{9,e}$, X.~C.~Ai$^{1}$, O.~Albayrak$^{5}$,
M.~Albrecht$^{4}$, D.~J.~Ambrose$^{44}$, A.~Amoroso$^{49A,49C}$,
F.~F.~An$^{1}$, Q.~An$^{46,a}$, J.~Z.~Bai$^{1}$
% ...

\center{(BESIII Collaboration)}\\

}
\address{
\vspace{0.0cm} {\it
$^{1}$ Institute of High Energy Physics, Beijing 100049, People's Republic of China\\
% ...
$^{4}$ Bochum Ruhr-University, D-44780 Bochum, Germany\\
$^{5}$ Carnegie Mellon University, Pittsburgh, Pennsylvania 15213, USA\\
% ...
$^{9}$ G.I. Budker Institute of Nuclear Physics SB RAS (BINP), Novosibirsk 630090, Russia\\
% ...
$^{44}$ University of Rochester, Rochester, New York 14627, USA\\
% ...
$^{46}$ University of Science and Technology of China, Hefei 230026, People's Republic of China\\
% ...
$^{49}$ (A)University of Turin, I-10125, Turin, Italy; (B)University of Eastern Piedmont, I-15121, Alessandria, Italy; (C)INFN, I-10125, Turin, Italy\\
% ...
$^{a}$ Also at State Key Laboratory of Particle Detection and Electronics, Beijing 100049, Hefei 230026, People's Republic of China\\
% ...
$^{e}$ Also at the Novosibirsk State University, Novosibirsk, 630090, Russia\\
% ...
}
}
\begin{abstract}
A measurement of the number of $J/\psi$ events collected with the BESIII
detector in 2009 and 2012 is performed using inclusive decays of the
$J/\psi$. The number of $J/\psi$ events taken in 2009
is recalculated to be $(223.7\pm 1.4)\times10^6$,
which is in good agreement with the previous measurement, but with significantly improved precision
due to improvements in the
BESIII software. The number of $J/\psi$ events taken in 2012
is determined to be $(1086.9\pm6.0)\times10^{6}$. In total, the number of
$J/\psi$ events collected with the BESIII detector is measured to be
$(1310.6\pm7.0)\times10^{6}$, where the uncertainty is dominated by
systematic effects and the statistical uncertainty is negligible.
\end{abstract}
\begin{multicols}{2}

\section{Introduction}
Studies of $J/\psi$ decays have provided a wealth of information
since the discovery of the $J/\psi$ in 1974~\cite{jsi1}\cite{jsi2}.
Decays of the $J/\psi$ offer a clean laboratory for light hadron spectroscopy,
provide an insight into decay mechanisms and help in distinguishing
between conventional hadronic states and exotic states.

A lot of important progress in light hadron spectroscopy
has been achieved based on a sample of $(225.3\pm
2.8)\times 10^6$ $J/\psi$ events collected by the BESIII experiment~\cite{bes3}
in 2009. To further comprehensively study the $J/\psi$ decay mechanism, investigate
the light hadron spectrum, and search for exotic states, \emph{e.g.}
glueballs, hybrids and multi-quark states,  an additional,
larger $J/\psi$ sample was collected in 2012.
A precise determination of the number of $J/\psi$ events is essential for
analyses based on these data samples.
With improvements in the BESIII software, particularly in Monte Carlo (MC) simulations
and the reconstruction of tracks in the main drift chamber (MDC),
it is possible to perform a more precise measurement
of the number of $J/\psi$ events taken in 2009 and 2012.
The relevant data samples used in this analysis are listed in
Table~\ref{samples}.

We implement the same method as that used in the previous study~\cite{njsi2009} to
determine the number of $J/\psi$ events. The advantage of this
approach is that the detection efficiency of inclusive $J/\psi$ decays can
be extracted directly from the data sample taken at the peak of the $\psi(3686)$.
This is useful because the correction factor of the detection efficiency
is less dependent on the MC model for
the inclusive $J/\psi$ decay
and therefore the systematic uncertainty can be reduced
significantly. The number of $J/\psi$ events, $N_{J/\psi}$, is calculated as
\begin{eqnarray}
\label{Nojpsi}
 N_{J/\psi}=\frac{N_\text{sel}-N_\text{bg}}{\epsilon_\text{trig} \times
\epsilon^{\psi(3686)}_\text{data} \times f_\text{cor}},
\end{eqnarray}
where $N_\text{sel}$ is the number of inclusive $J/\psi$
events selected from the $J/\psi$ data; $N_\text{bg}$ is the number of
background events estimated with continuum data taken at $\sqrt{s} = 3.08\,\text{GeV}$;
$\epsilon_\text{trig}$ is the trigger efficiency; $\epsilon^{\psi(3686)}_\text{data}$ is the
inclusive $J/\psi$ detection efficiency determined experimentally
using the $J/\psi$ sample from the reaction $\psi(3686) \rightarrow \pi^+ \pi^-
J/\psi$. $f_\text{cor}$ is a correction factor that
accounts for the difference in the detection efficiency between the $J/\psi$ events produced at
rest and those produced in $\psi(3686)\rightarrow \pi^+ \pi^- J/\psi$.
$f_\text{cor}$ is expected to be unity approximately,  and is determined
by the MC simulation sample with
\begin{eqnarray}
\label{Fcor}
 f_\text{cor} = \frac {\epsilon^{J/\psi}_\text{MC}}  {\epsilon^{\psi(3686)}_\text{MC}},
\end{eqnarray}
where $\epsilon^{J/\psi}_\text{MC}$ is the detection efficiency of
inclusive $J/\psi$ events determined from
the MC sample of $J/\psi$ events produced directly in the electron--positron
collision, and $\epsilon^{\psi(3686)}_\text{MC}$ is that from the  MC sample of
$\psi(3686)\to \pi^+\pi^- J/\psi ~(J/\psi\rightarrow inclusive)$ events.
In MC simulation, the $J/\psi$ and $\psi(3686)$ resonances are simulated with
KKMC~\cite{kkmc}. The known decay modes of the $J/\psi$ and
$\psi(3686)$ are generated by EVTGEN~\cite{evtgen,djl} with branching fractions taken from the
Review of Particle Physics~\cite{PDG}, while the remaining decays are generated according to the
LUNDCHARM model~\cite{LUND,LUND2}.  All of the MC events are fed into a
GEANT4-based~\cite{geant4} simulation package, which takes into account the detector geometry and response.


\begin{center}
\tabcaption{\label{samples}Data samples used in the determination of the
number of $J/\psi$ events collected in 2009 and 2012.}
\footnotesize
\begin{tabular*}{75mm}[htbp]{c@{\extracolsep{\fill}}cccc}
\toprule Data set & $\sqrt{s}$ &$\calL_\text{online}$ & Date(duration) \\
 &&&(MM/DD/YYYY)\\\hline
    $J/\psi$  & 3.097 GeV& $323$pb$^{-1}$ & 4/10/2012--5/22/2012 \\
    QED1  & 3.08 GeV& $13$pb$^{-1}$ & 4/8/2012 \\
    QED2  & 3.08 GeV& $17$pb$^{-1}$ & 5/23/2012--5/24/2012 \\
    $\psi(3686)$  & 3.686 GeV& $7.5$pb$^{-1}$ & 5/26/2012 \\\hline
    $J/\psi$  & 3.097 GeV& $82$pb$^{-1}$ & 6/12/2009--7/28/2009 \\
    QED  & 3.08 GeV& $0.3$pb$^{-1}$ & 6/19/2009 \\
    $\psi(3686)$  & 3.686 GeV& $150$pb$^{-1}$ & 3/7/2009--4/14/2009 \\\hline
\bottomrule
\end{tabular*}
\end{center}

\section{\boldmath Inclusive $J/\psi$ selection criteria}
\label{hadsel}
To distinguish the  inclusive $J/\psi$ decays from
Quantum Electro-Dynamics (QED)  processes (\emph{i.e.} Bhabha and dimuon
events) and background events from cosmic rays and beam-gas
interactions, a series of selection criteria are applied to the candidate events.
The charged tracks are required to be detected in the MDC
 within a polar angle range of $|\cos\theta| < 0.93$, and to have a momentum
of $p<2.0\;\text{GeV}/c$. Each track is  required to originate from
the interaction region by restricting the distance of closest
approach to the run-dependent interaction point in the radial direction,
 $V_{r}<1\;\text{cm}$, and in the beam direction, $|V_z|<15\;\text{cm}$.
For photon clusters in the electromagnetic calorimeter (EMC), the deposited
energy is required be greater than 25 (50) MeV for the barrel (endcap) region
of $|\cos\theta| < 0.83$ ($0.86< |\cos\theta| < 0.93$). In addition, the
EMC cluster timing $T$ must satisfy 0 $ <T\le$ 700 ns, which
is used to suppress electronics noise and energy deposits unrelated to the event.

The candidate event must contain at least two charged tracks. The visible energy $E_\text{vis}$,
defined as the sum of charged particle energies computed
from the track momenta by assuming a pion mass and from the neutral shower
energies deposited in the EMC, is required to be greater than 1.0 GeV.
A comparison of the $E_\text{vis}$ distribution between the $J/\psi$ data, the data
taken at $\sqrt{s} = 3.08$ GeV, and the inclusive $J/\psi$ MC sample is illustrated
in Fig.~\ref{evis}. The requirement $E_\text{vis} > 1.0~\text{GeV}$ removes one third of
the background events while retaining $99.4\%$ of the signal events.

\begin{center}
\includegraphics[width=8.cm,height=5.7cm]{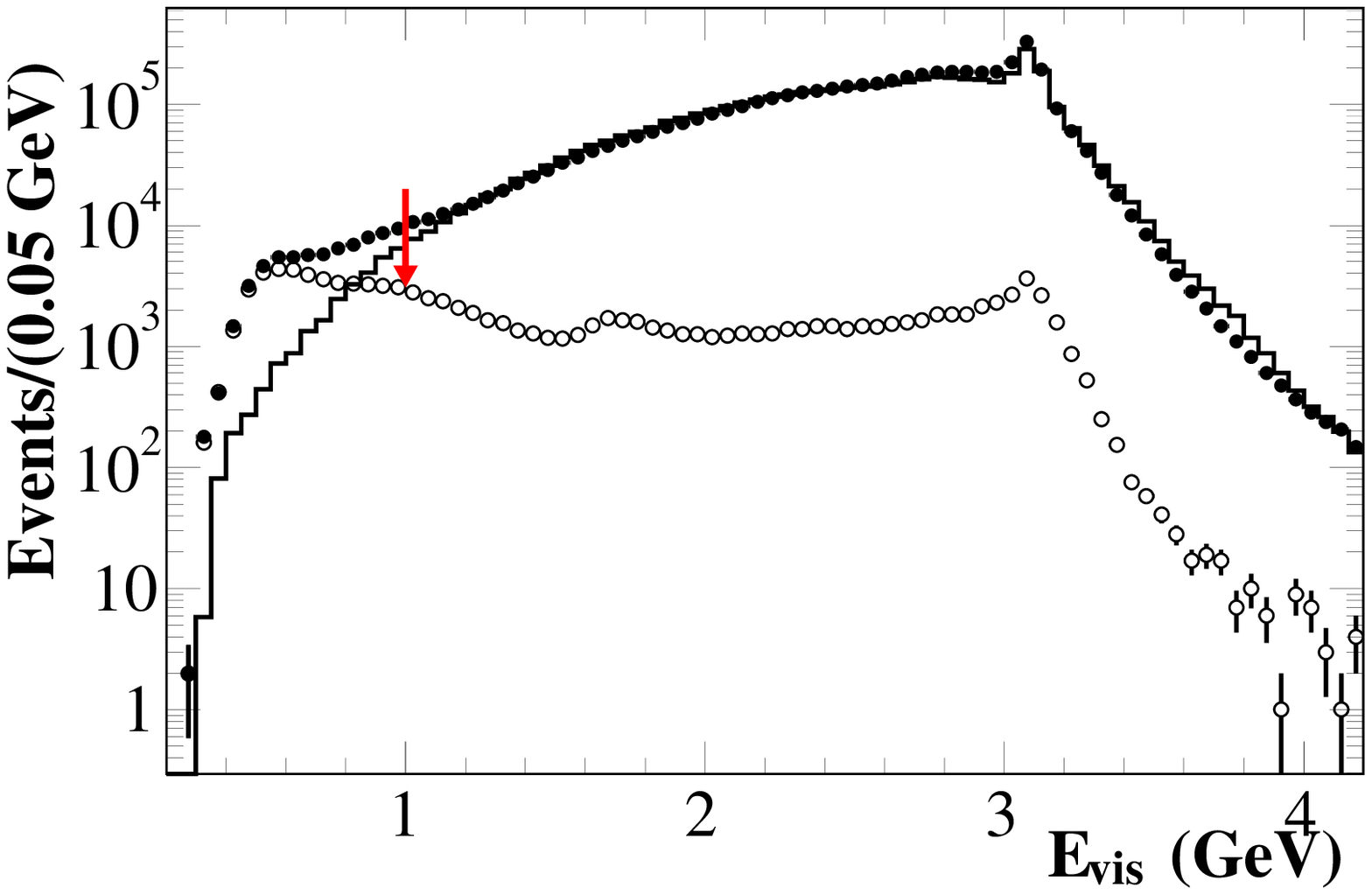}
\figcaption{\label{evis} Distributions of the visible energy $E_\text{vis}$ for $J/\psi$
data (dots with error bars), continuum data
at $\sqrt{s}=3.08$ GeV
(open circles with error bars)
and MC simulation of inclusive $J/\psi$ events
(histogram). The arrow indicates the minimum $E_\text{vis}$ required to
select inclusive events.  }
\end{center}
\vspace{-0.55cm}

Since Bhabha ($e^+e^-\rightarrow e^+e^-$) and dimuon
($e^+e^-\rightarrow\mu^+\mu^-$) events are two-body decays,
each charged track carries a unique energy, close to
half of the center-of-mass energy. Therefore, for events with
only two charged tracks, we require that the momentum of each charged
track is less than 1.5 GeV/$c$ in order to remove
Bhabha and dimuon events. This requirement is  depicted by the solid lines in
the scatter plot of the momenta of the two charged tracks (Fig.~\ref{pcutn}).
The Bhabha events are characterized by a significant peak around 1.5 GeV
in the distribution of energy deposited in the EMC, shown in Fig.~\ref{etrk}.
Hence an additional requirement that the energy deposited  in the EMC for each
charged track is less than 1 GeV is applied to further reject the Bhabha events.
\begin{center}
\includegraphics[width=8.0cm,height=5.7cm]{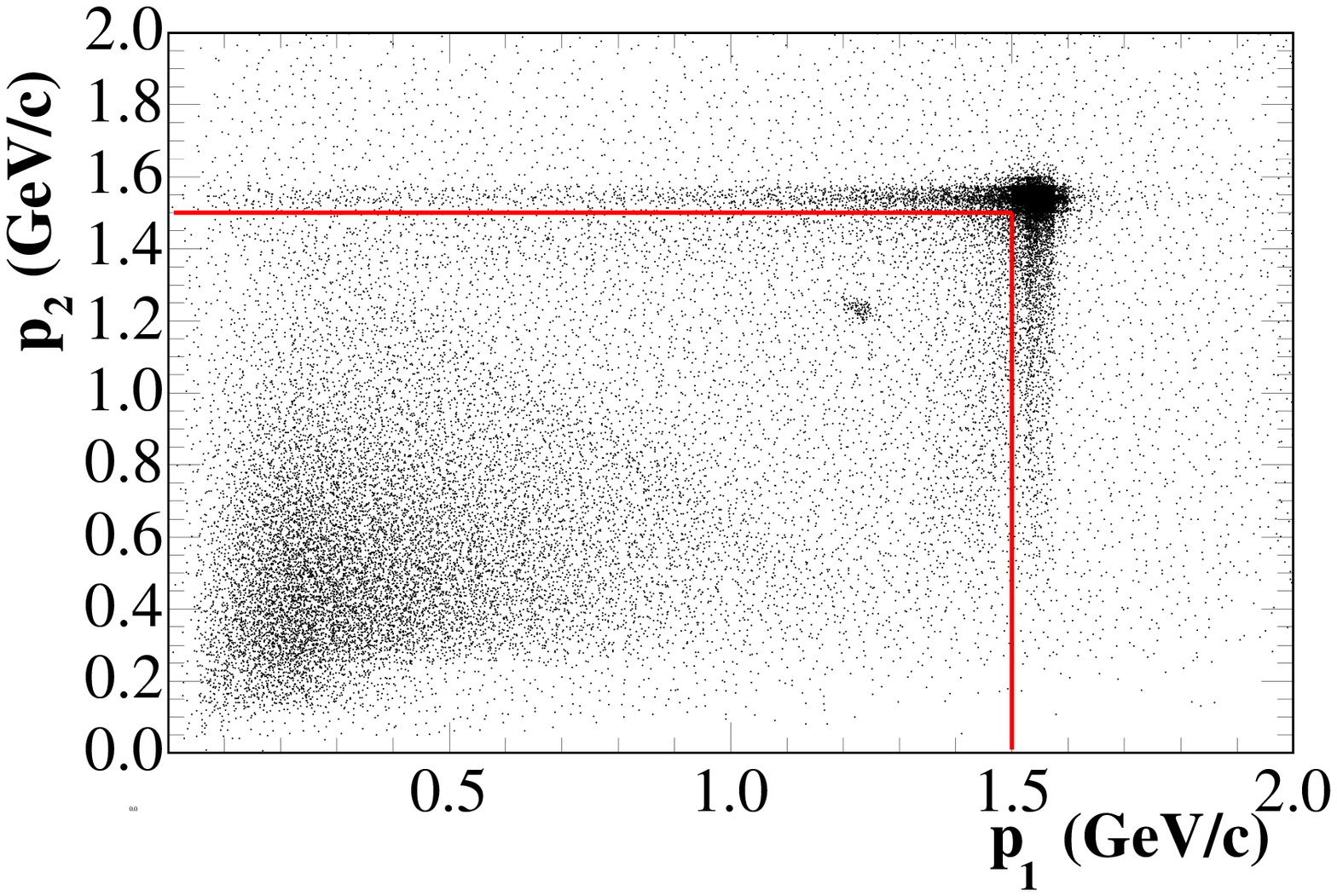}
\figcaption{\label{pcutn} Scatter plot of the momenta of the
charged tracks for 2-prong events in data. The cluster around 1.55 GeV/$c$
corresponds to the contribution from lepton pairs and the cluster at 1.23
GeV/$c$ comes from $J/\psi \rightarrow p\bar{p}$.  Most of lepton pairs are removed
with the requirements on the two charged tracks, $p_1<1.5$ GeV/$c$ and
$p_2<1.5$ GeV/$c$, as indicated by the solid lines.}
\end{center}
\begin{center}
\includegraphics[width=8.0cm,height=5.7 cm]{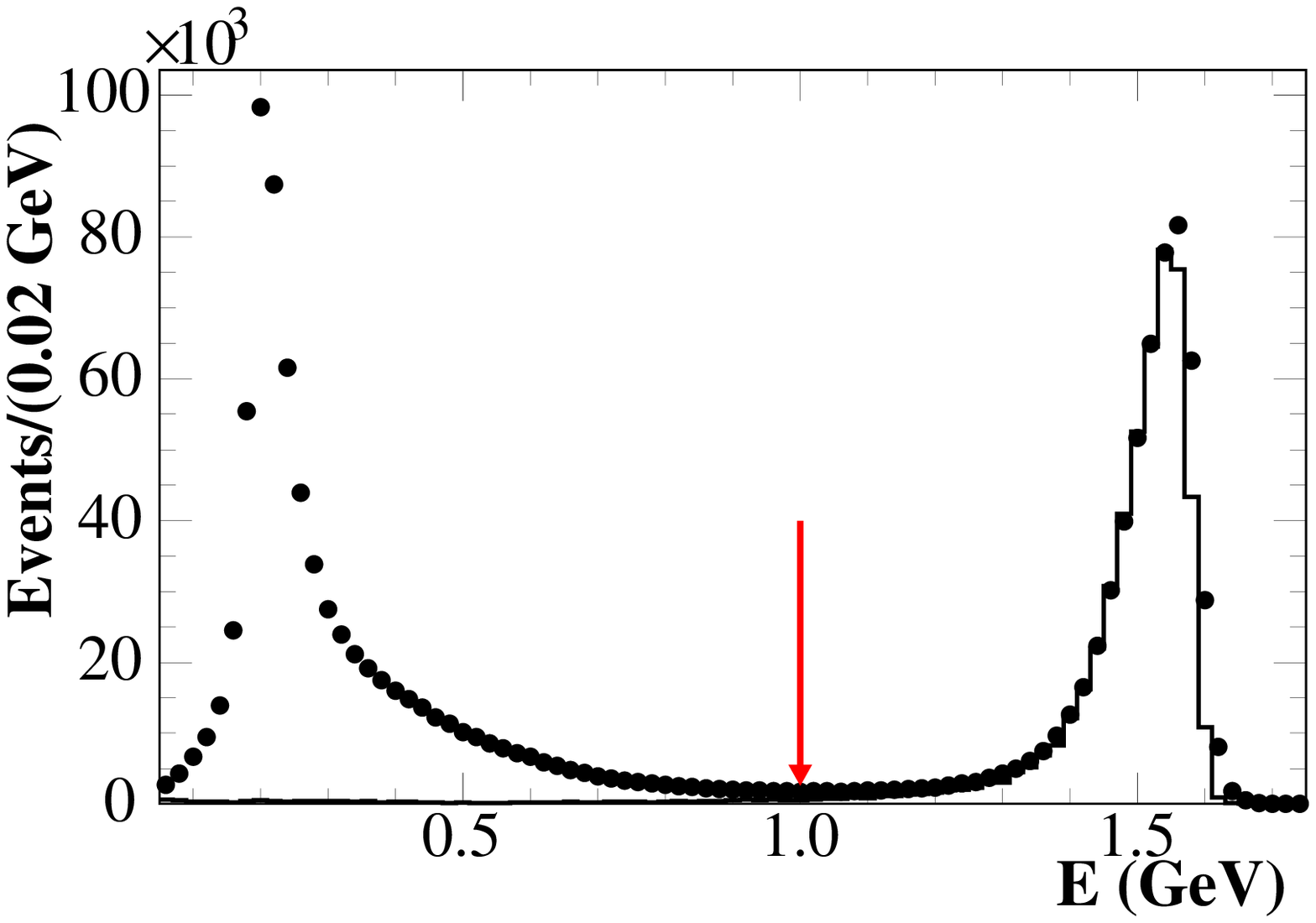}
\figcaption{\label{etrk} Distributions of deposited energy in the
  EMC for the charged tracks of 2-prong events for $J/\psi$ data (dots
  with error bars) and for the combined, normalized MC simulations of
  $e^+e^-\rightarrow e^+e^-(\gamma)$ and $J/\psi\rightarrow e^+e^-(\gamma)$
  (histogram).}
\end{center}

After the above requirements, $N_\text{sel} = (854.60 \pm 0.03)\times 10^{6}$ candidate
events are selected from the $J/\psi$ data taken in 2012. The
distributions of the track parameters of closest approach in the beam line and
radial directions ($V_z$ and $V_r$), the polar angle ($\cos\theta$),
and the total energy deposited in the EMC ($E_\text{EMC}$) after subtracting
background events  estimated with the continuum data taken at $\sqrt{s} = 3.08$
GeV (see Sec.~{\ref{bkg}} for details) are shown in Fig.~\ref{vz0}.
Reasonable agreement between the data and MC samples is observed.
The multiplicity of charged tracks ($N_\text{good}$) is
shown in Fig.~\ref{ngood}, where the MC sample generated according to the LUNDCHARM model
agrees very well with the data while the MC sample generated without the LUNDCHARM
model deviates from the data. The effect of this discrepancy on the
determination of the number of $J/\psi$ events is small, as described
in Sec.~{\ref{syst}}.

\begin{center}
\includegraphics[width=3.9cm,height=3.2cm]{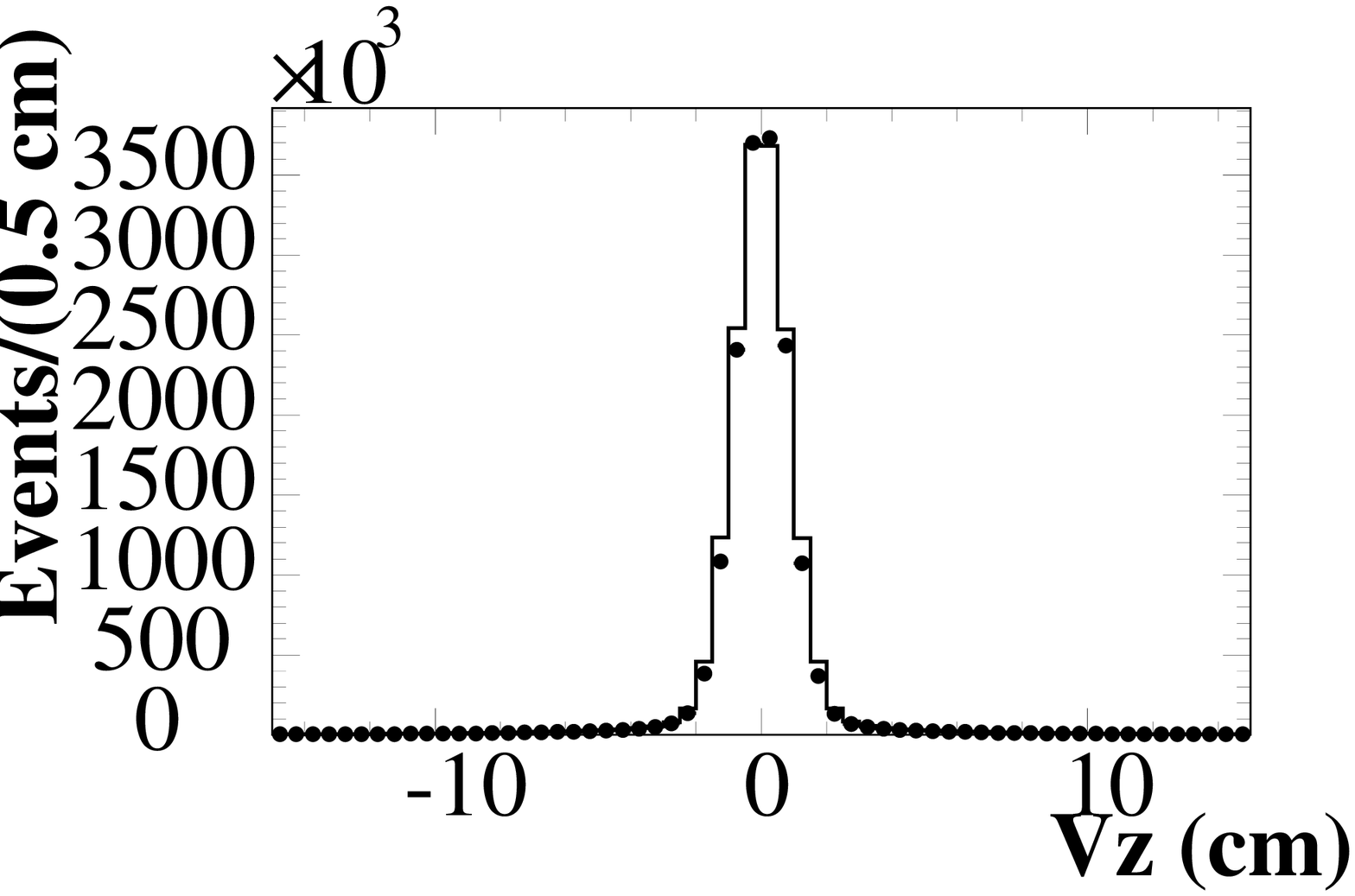}
\includegraphics[width=3.9cm,height=3.2cm]{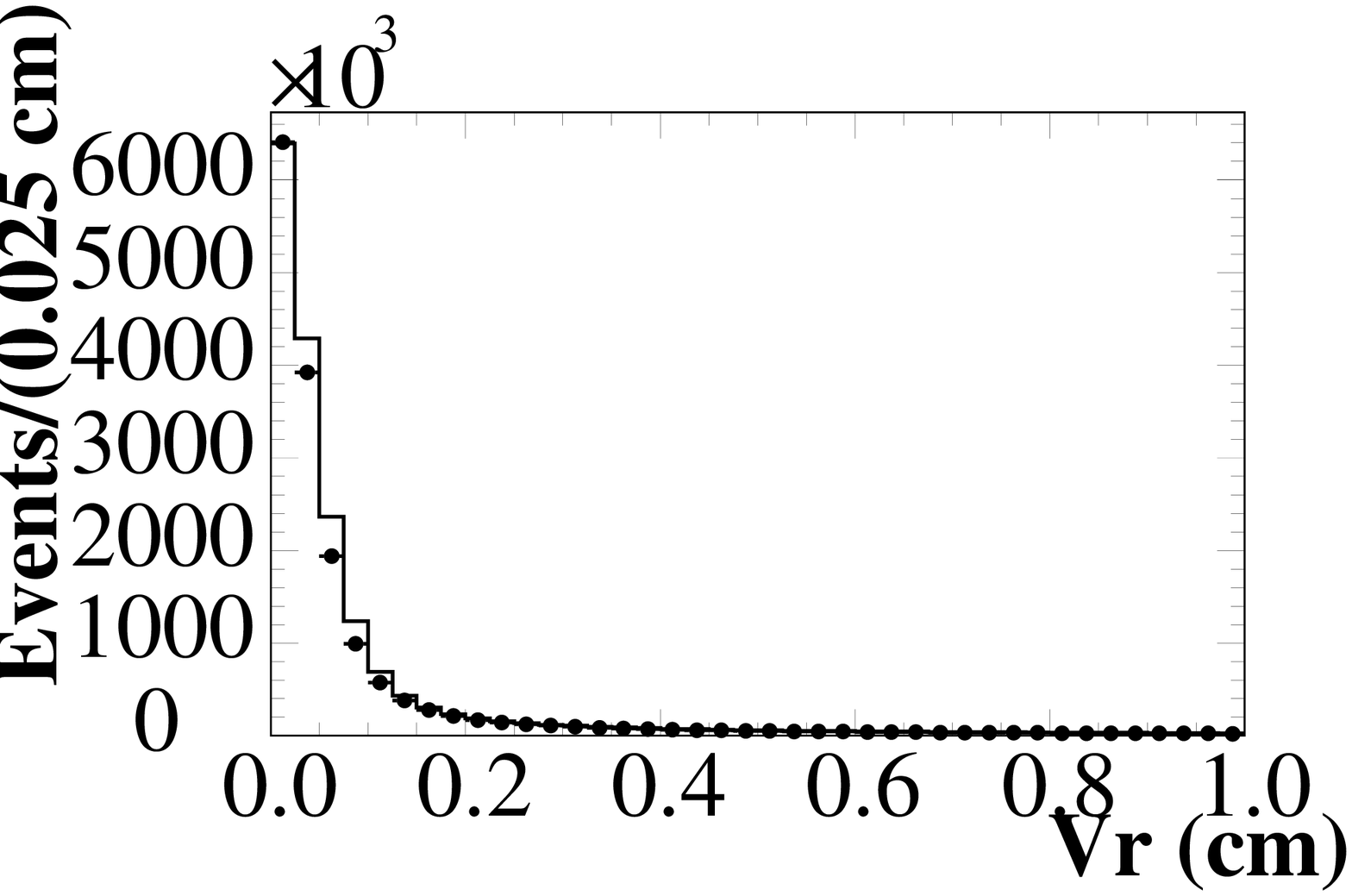}
\put(-143,60){(a)}
\put(-30,60){(b)}\\
\includegraphics[width=3.9cm,height=3.2cm]{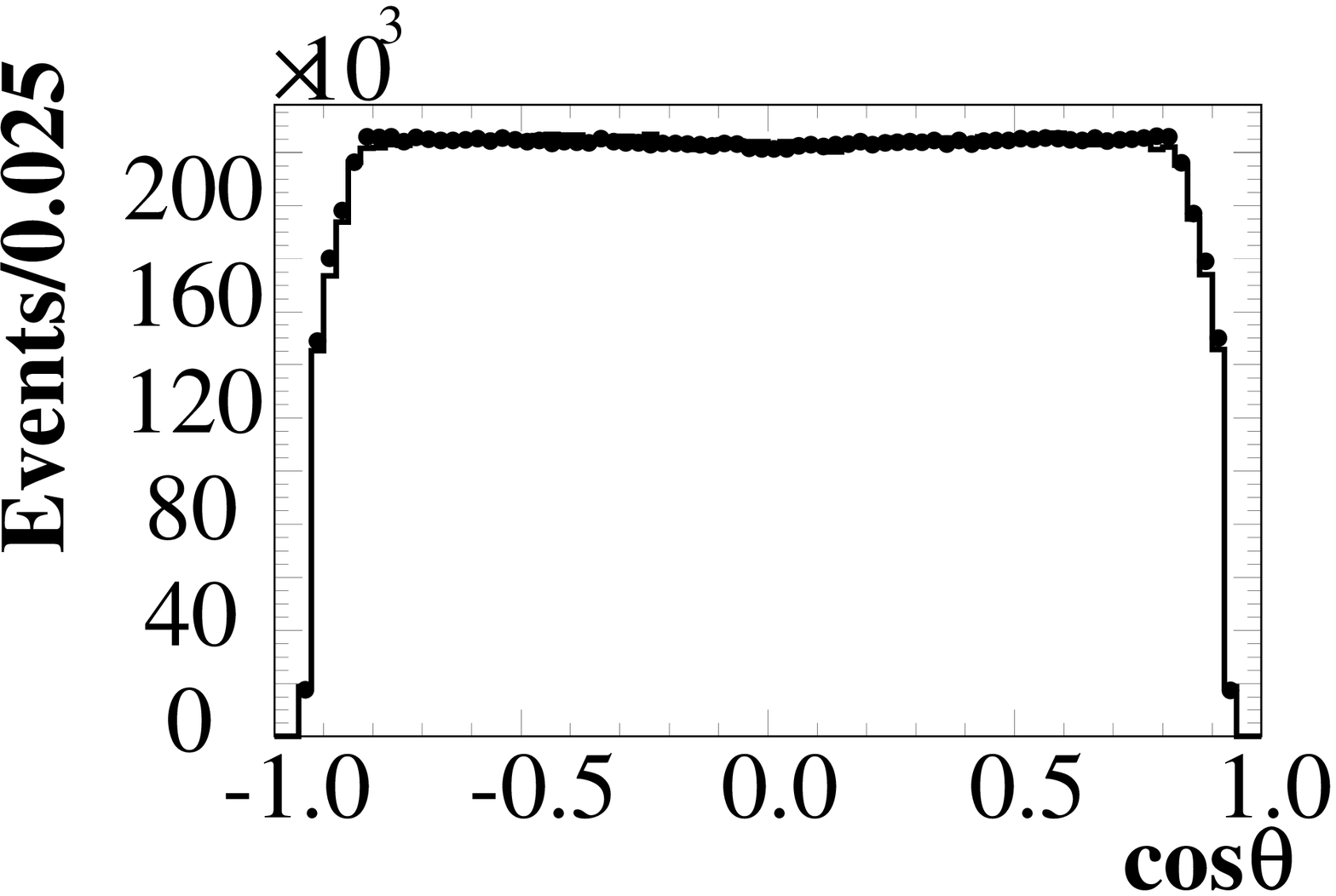}
\includegraphics[width=3.9cm,height=3.2cm]{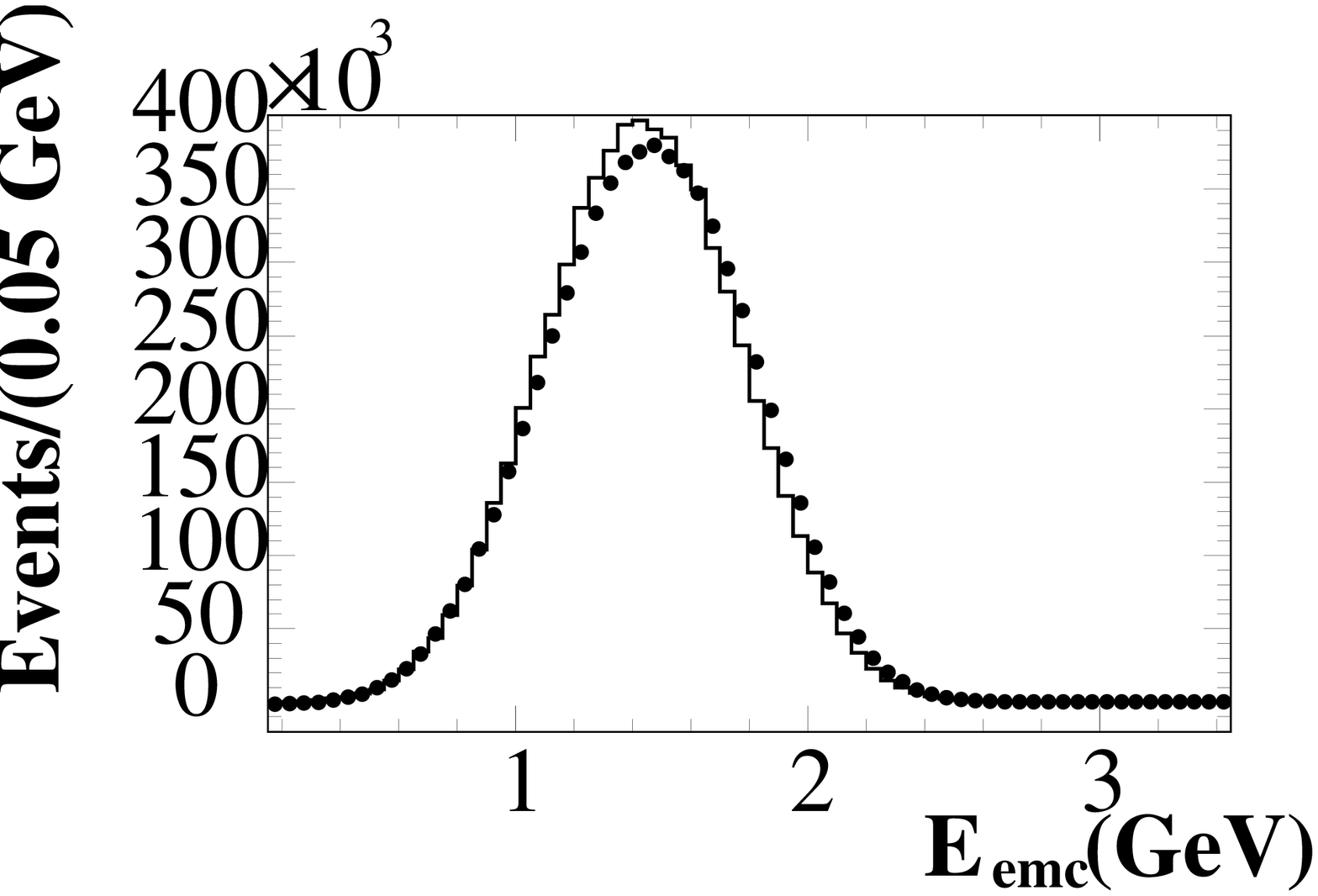}
\put(-143,60){(c)}
\put(-30,60){(d)}
\figcaption{\label{vz0} Comparison of distributions between $J/\psi$ data (dots with error
bars) and MC simulation of inclusive $J/\psi$ (histogram): (a) $V_z$,
(b) $V_r$, (c) $\cos\theta$
of charged tracks, (d) total energy deposited in the EMC.}
\end{center}

\begin{center}
\includegraphics[width=8.0cm,height=5.7cm]{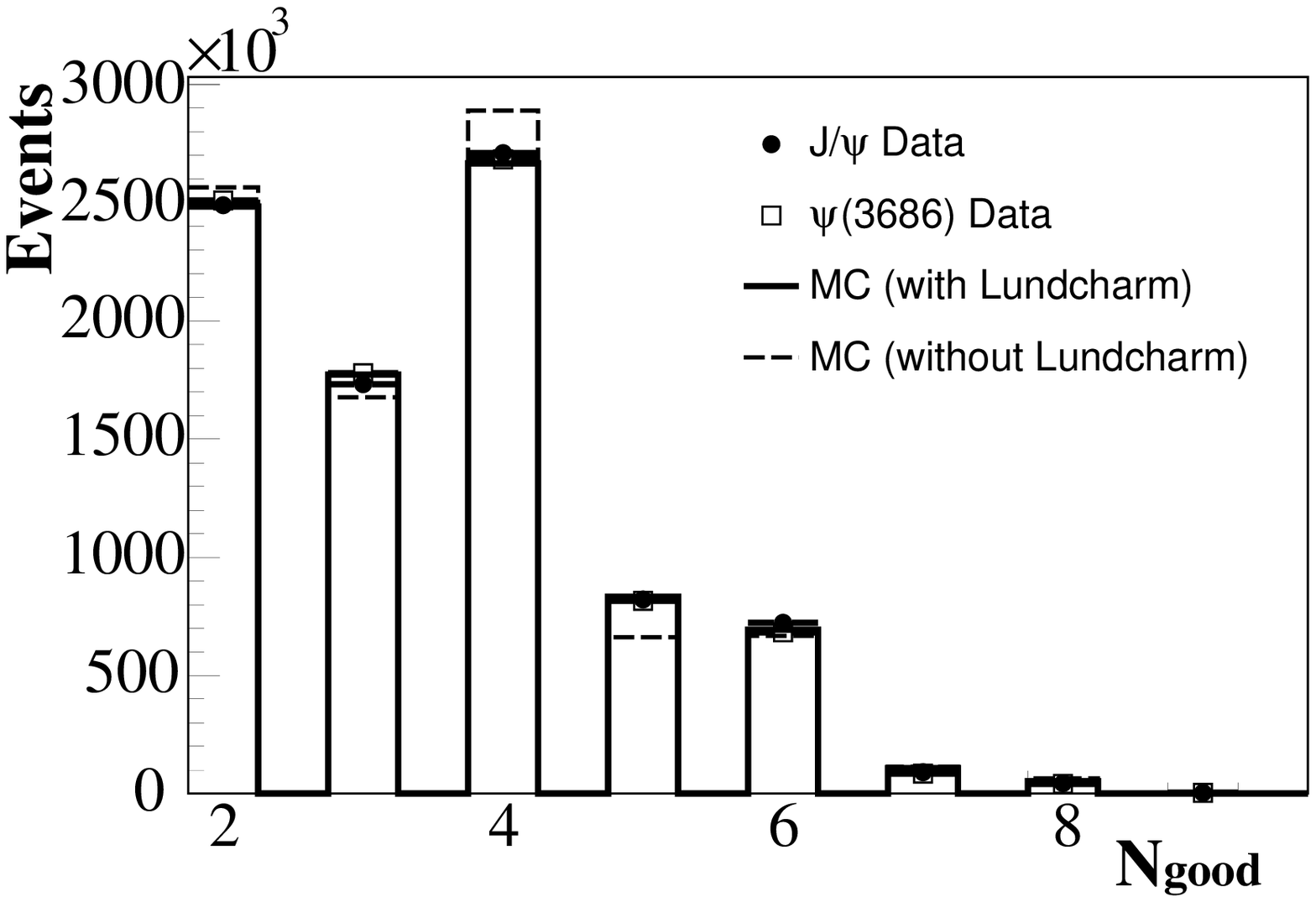}
\figcaption{\label{ngood} Distributions of the reconstructed charged track
multiplicity of inclusive $J/\psi$ events for $J/\psi$ data
(dots with error bars) and $\psi(3686)$ data (squares with error bars) and
MC simulation generated with and without the LUNDCHARM model (solid and
dashed histograms, respectively). }
\end{center}

\section{Background analysis}
\label{bkg}

In this analysis, the data samples taken at $\sqrt{s}=3.08$ GeV and
in close chronological order to the $J/\psi$ sample are used to estimate the
background due to QED processes,  beam-gas interactions and cosmic rays. To
normalize the selected background events to the $J/\psi$ data, the integrated
luminosity for the data samples taken at the $J/\psi$ peak and at $\sqrt{s}=3.08$ GeV
 are determined using the precess $e^+e^- \rightarrow \gamma
\gamma$, respectively.

To determine the integrated luminosity, the candidate events
$e^+e^-\rightarrow \gamma\gamma$ are selected by requiring at least two
showers in the EMC. It is further required that the energy of the second most energetic
shower is between 1.2 and 1.6 GeV and that the polar angles of the two
showers are in the range $|\cos\theta | < 0.8$. The number of signal events is determined from
the number of events in the signal region $|\Delta \phi| < 2.5^\circ$ and the background
is estimated from those in the sideband region $2.5 < |\Delta \phi| < 5^\circ$,
where $\Delta \phi = |\phi_{\gamma1} - \phi_{\gamma2}| - 180^\circ $
and $\phi_{\gamma1/2}$ is the azimuthal angle of the photon.
Taking into account the detector efficiency obtained from the MC simulation
and the cross section of the QED process  $e^+e^- \rightarrow
\gamma\gamma$, the integrated luminosities of the $J/\psi$ data sample and
the sample taken at $\sqrt{s}=3.08$ GeV taken in 2012 are determined to be $315.02 \pm 0.14
\;\text{pb}^{-1}$ and $30.84 \pm 0.04\;\text{pb}^{-1}$, respectively,
where the errors are statistical only.

After applying the same selection criteria as for the $J/\psi$ data,
$N_{3.08}=1,440,376\pm1,200$ events are selected from the continuum data taken
at $\sqrt{s} = 3.08$ GeV. Assuming the same detection efficiency at
$\sqrt{s} = 3.08$ GeV as for the $J/\psi$ peak and taking into account the
energy-dependent cross section of the QED processes, the number of background
events for the $J/\psi$ sample, $N_\text{bg}$, is estimated to be
\begin{eqnarray}
\label{Nbg} N_\text{bg}=N_{3.08}\times
{\frac{\calL_{J/\psi}}{\calL_{3.08}} \times
\frac{s_{3.08}}{s_{J/\psi}}}=(14.55 \pm 0.02)\times 10^{6} ,
\end{eqnarray}
where $\calL_{J/\psi}$ and $\calL_{3.08}$ are
the integrated luminosities for the $J/\psi$ data sample and
the data sample taken at $\sqrt{s}=3.08$ GeV,
respectively, and $s_{J/\psi}$ and $s_{3.08}$ are the
corresponding squares of the center-of-mass energies.
The background is calculated to be 1.7\% of
the selected inclusive $J/\psi$  events taken in 2012.

According to the studies of the MC sample and the $V_z$ distribution, the QED
background fraction is found to be about 1.5\% of the total data.
It is known that the beam status for the data taken in 2009 was worse and the background is much
higher than for the 2012 sample.
With the same method, the total background (including the QED contribution) for the 2009 sample
is estimated to be 3.7\%.

\section{Determination of the detection efficiency and correction factor}
\label{eff-corr-factor}
In the previous study, the detection efficiency was determined using a MC simulation of the reaction
$J/\psi \to \text{inclusive}$, assuming that both the physics process
of the inclusive $J/\psi$ decay and the detector response were simulated well.
In this analysis, to reduce the uncertainty related to the discrepancy
between the MC simulation and the data, the detection efficiency is determined
experimentally using a sample of $J/\psi$ events from the
reaction $\psi(3686) \rightarrow \pi^+\pi^- J/\psi$.
To ensure that the beam conditions and detector status are similar to those of the sample collected at the
$J/\psi$ peak, a dedicated $\psi(3686)$ sample taken on May 26, 2012 is used for this study.

To select $\psi(3686) \rightarrow \pi^+\pi^- J/\psi$ events, there
must be at least two soft pions
with opposite charge in the MDC within the polar angle range $|\cos\theta| <
0.93$, having $V_{r}<1\;\text{cm}$ and $|V_{z}|<15\;\text{cm}$, and momenta less
than $0.4\;\text{GeV}/c$. No further selection criteria on the remaining charged tracks
or showers are required. The distribution of the invariant mass recoiling against all
possible soft $\pi^+\pi^-$ pairs is shown in
Fig.~\ref{fitdat} (a). A prominent peak around $3.1\;\text{GeV}/c^2$, corresponding
to the decay of $\psi(3686) \rightarrow \pi^+\pi^- J/\psi$,
$J/\psi\rightarrow inclusive$, is observed over a smooth
background. The total number of inclusive $J/\psi$ events,
$N_\text{inc} = (1147.8\pm 1.9)\times 10^{3}$, is obtained by fitting
a double-Gaussian function for the $J/\psi$ signal plus a second order
Chebychev polynomial for the background to the
$\pi^+$$\pi^-$ recoil mass spectrum.

To measure the detection efficiency of inclusive $J/\psi$ events, the same
selection criteria as described in Sec.~{\ref{hadsel}} are applied to the remaining
charged tracks and showers at the event level. The distribution of the
invariant mass
recoiling against $\pi^+\pi^-$ for the remaining events is shown in
Fig.~\ref{fitdat} (b); it is fitted with the same function described above.
The number of selected inclusive $J/\psi$ events, $N_\text{inc}^\text{sel}$,
is determined to be $(877.6\pm 1.7)\times 10^{3}$. The detection
efficiency of inclusive $J/\psi$ events, $\epsilon^{\psi(3686)}_\text{data} =
(76.46\pm0.07)\%$, is calculated by the ratio of the number of
inclusive $J/\psi$ events with and without the inclusive $J/\psi$ event
selection criteria applied.

Since the $J/\psi$  particle in the decay $\psi(3686) \rightarrow \pi^+\pi^-
J/\psi$ is not at rest, a correction factor, defined in Eq.~(\ref{Fcor}), is used to take into account
the kinematical effect on the detection efficiency of the inclusive $J/\psi$ event
selection. Two large statistics, inclusive $\psi(3686)$ and $J/\psi$ MC samples are produced
and are subjected to the same selection criteria as the data samples.
The detection efficiency of
inclusive $J/\psi$ events are determined to be $\epsilon^{\psi(3686)}_\text{MC} =
(75.76\pm 0.06)$\%, and $\epsilon^{J/\psi}_\text{MC}=(76.58 \pm 0.04)$\%
for the two inclusive MC samples, respectively.
The correction factor $f_\text{cor}$ for the detection efficiency is therefore taken as

\begin{eqnarray}
\label{Fcorr} f_\text{cor} = \frac {\epsilon^{J/\psi}_\text{MC}}
{\epsilon^{\psi(3686)}_\text{MC}}=1.0109 \pm 0.0009.
\end{eqnarray}

\begin{center}
\includegraphics[width=8.0cm,height=5.7cm]{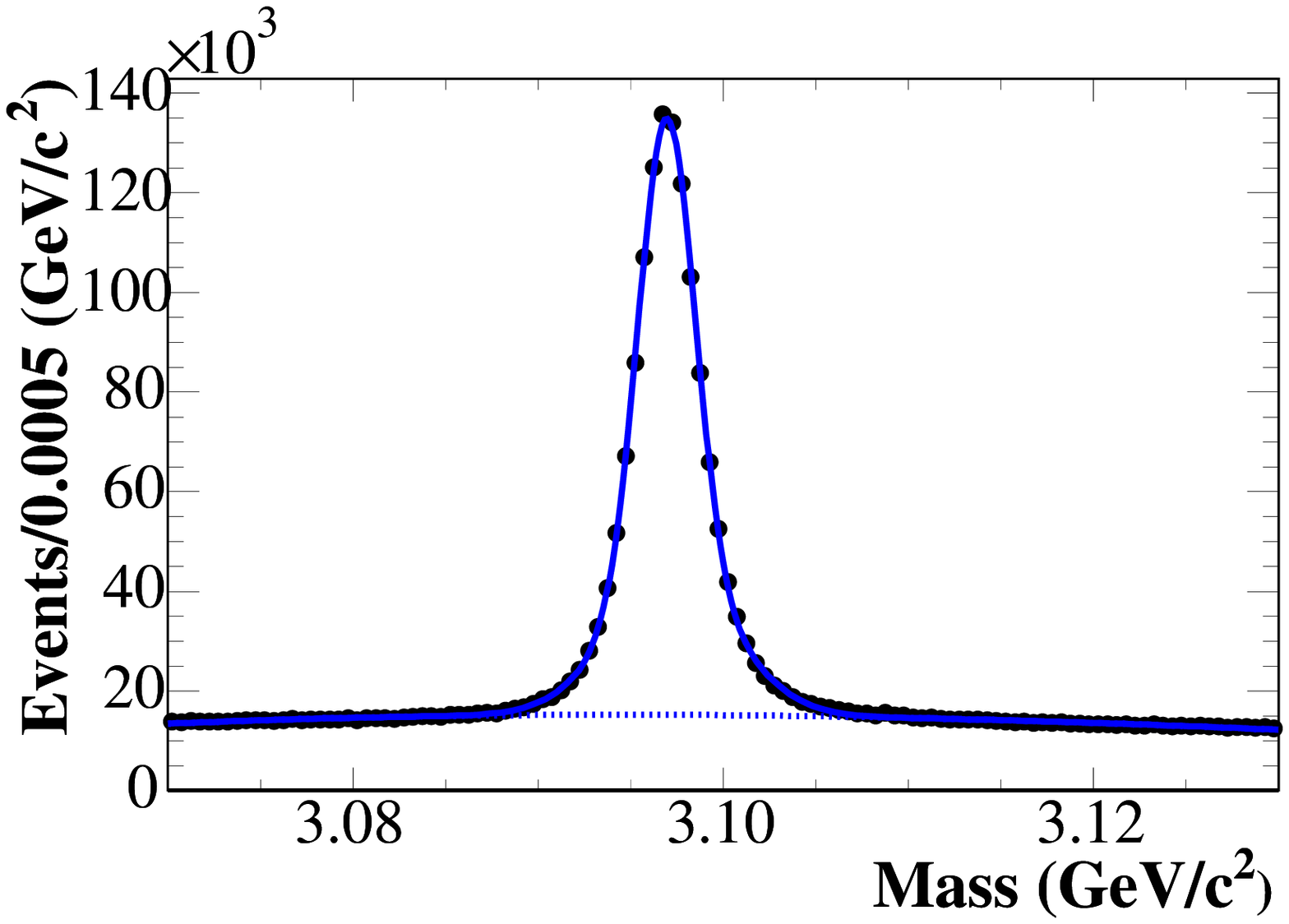}
 \put(-150,110){(a)}\\
\includegraphics[width=8.0cm,height=5.7cm]{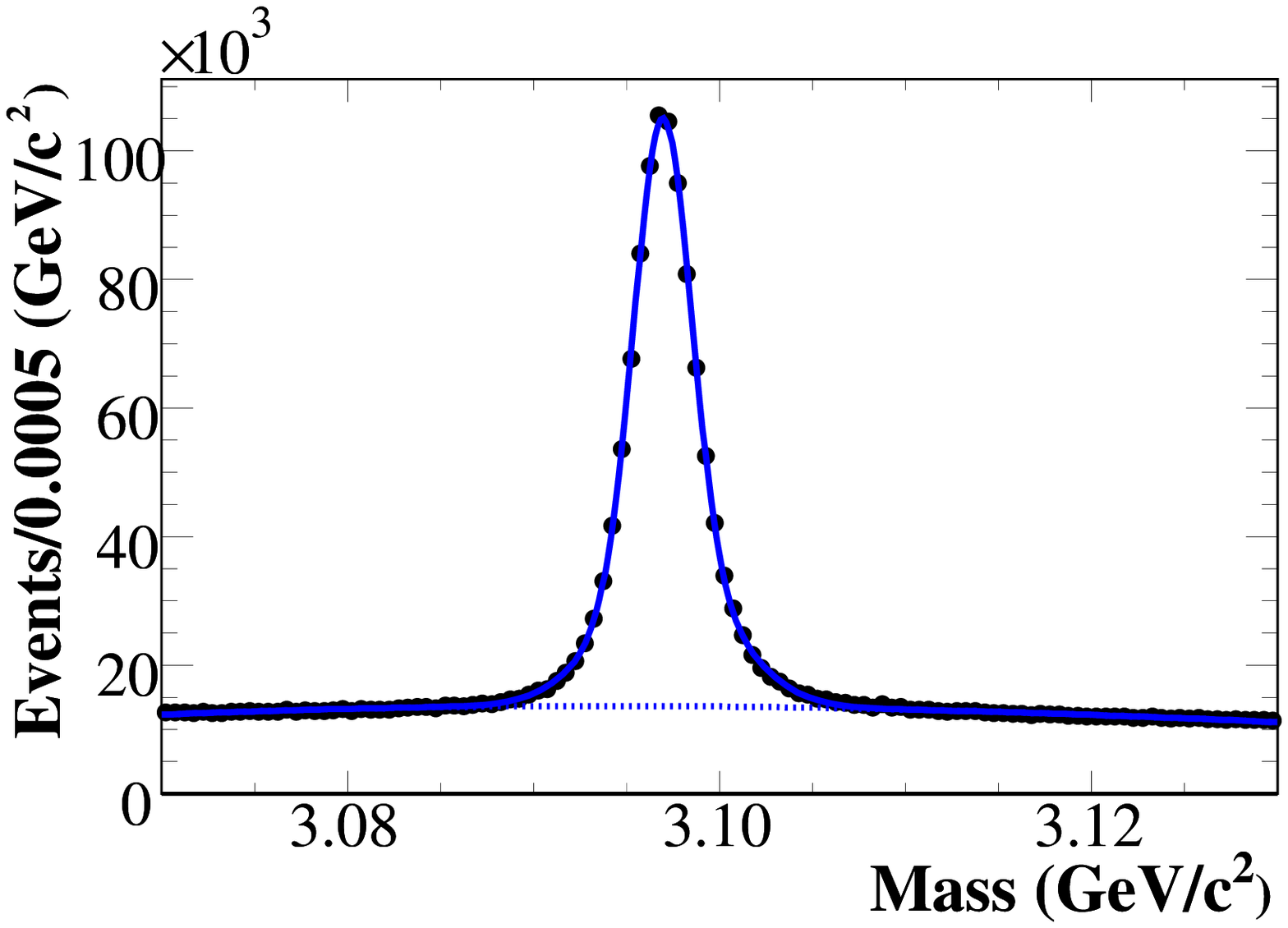}
 \put(-150,110){(b)}
\figcaption{\label{fitdat}Invariant mass recoiling against selected
$\pi^+\pi^-$ pairs for the $\psi(3686)$ data sample. The curves are the
results of the fit described in the text:
(a) for the sample with  soft pion selection criteria applied, and
(b) for the sample with the addition of the inclusive $J/\psi$ event selection criteria applied.}
\end{center}

\section{\boldmath The number of $J/\psi$ events}
Using Eq.~(\ref{Nojpsi}), the number of $J/\psi$ events collected in 2012 is calculated to
be $(1086.9 \pm 0.04) \times 10^6$. The values used in this calculation are summarized in
Table~\ref{Nformula}. The trigger efficiency of the BESIII detector is 100\%, based on the
study of various reactions~\cite{trig}. With the same procedure,
the  number of $J/\psi$ events taken in 2009 is determined to be $(223.72
\pm 0.01) \times 10^6$.
Here, the statistical uncertainty is from the number of $J/\psi$ events only,
while the statistical fluctuation of $N_\text{bg}$ is taken into account as
part of the systematic uncertainty (see Sec.~{\ref{bkgerror}}). The systematic
uncertainties from different sources are discussed in detail in
Sec.~{\ref{syst}}.

\begin{center}
\tabcaption{\label{Nformula}Summary of the values used in the calculation
and the resulting number of $J/\psi$ events.}
\footnotesize

\begin{tabular*}{75mm}{c@{\extracolsep{\fill}}cc}
\toprule Item & 2012&2009\\ \hline
$N_\text{sel}$ &$(854.60\pm0.03)\times10^6$   &$(179.63 \pm 0.01) \times 10^{6}$ \\
$N_\text{bg}$ & $(14.55\pm0.02)\times10^6$    &$(6.58 \pm 0.04) $$\times 10^{6}$ \\
$\epsilon_\text{trig}$ & 1.00&1.00 \\
$\epsilon^{\psi(3686)}_\text{data}$ & $0.7646\pm0.0007$   &$0.7655 \pm 0.0001$  \\
$\epsilon^{\psi(3686)}_\text{MC}$ &$0.7576\pm0.0006$ & $0.7581 \pm 0.0005$ \\
$\epsilon^{J/\psi}_\text{MC}$ & $0.7658\pm0.0004$ &$0.7660 \pm 0.0004  $ \\
$f_\text{cor}$ & $1.0109\pm0.0009$ &$1.0105 \pm 0.0009$ \\\midrule
$N_{J/\psi}$ & $(1086.90\pm0.04)\times10^6$ &$(223.72 \pm 0.01) \times10^6$  \\
\bottomrule
\end{tabular*}
\end{center}

\section{Systematic uncertainty}
\label{syst}
The sources of systematic uncertainty and their corresponding
contributions
are summarized in Table~\ref{TABSYS}, and are discussed in detail below.

\subsection {MC model uncertainty}
In the measurement of the number of $J/\psi$ events, only the efficiency
correction factor, $f_\text{cor}$, is dependent on the MC simulation.
To evaluate the uncertainty due to the MC model,
we generate a set of MC samples without the LUNDCHARM model
and compare the correction factor determined using these samples to its nominal value.
According to the
distributions of the charged track multiplicity shown in Fig.~\ref{ngood},
the MC simulation without the LUNDCHARM model poorly
describes the data, which means this method will overestimate
the systematic uncertainty. The studies show that the correction factor
has a slight dependence on the MC mode of inclusive $J/\psi$ decays. To be
conservative, the change
in the correction factor, 0.42\% (0.36\%), is taken as the systematic
uncertainty due to the MC model on the number of $J/\psi$ events taken in 2012 (2009).

\subsection {Track reconstruction efficiency}
According to studies of the track reconstruction efficiency,
the difference in track reconstruction efficiencies between the MC and data samples of
$J/\psi$ decays is less than $1$\% for each charged track.

In the analysis, the $\psi(3686)$ data sample used to determine the detection
efficiency is taken in close chronological order to the $J/\psi$ sample.
The consistency of track reconstruction efficiency between
the MC and data samples in $\psi(3686)$ decays is assumed to be exactly the same as
that in $J/\psi$ decays.  Therefore the track reconstruction
efficiencies in both $J/\psi$ and $\psi(3686)$ MC samples are varied by
$-1\%$ to evaluate the uncertainty due to the MDC tracking.
As expected, the change in the correction factor is very small, $0.03\%$, and
this is taken as a systematic uncertainty.

In the determination of the number of $J/\psi$ events taken in 2009, the $J/\psi$ and
$\psi(3686)$ data samples were collected at different times, which may lead to slight
differences in the tracking efficiency between the two data sets due to the imperfect description of detector
performance and response in the MC simulation.
To estimate the corresponding systematic uncertainty, we adjust the track
reconstruction efficiency by $-0.5\%$ in the $J/\psi$ MC sample, keeping it
unchanged for the $\psi(3686)$ MC sample. The resulting change in the
correction factor, $0.30\%$, is
taken as a systematic uncertainty on the number of $J/\psi$ events in 2009.

\subsection{\boldmath Fit to the $J/\psi$ peak}
In this measurement, the selection efficiency of inclusive $J/\psi$ events is
estimated
experimentally  with the $\psi(3686)$ data sample ($\psi(3686) \rightarrow
\pi^+\pi^- J/\psi$), and
the yield of $J/\psi$ events used in the efficiency calculation is determined by a fit to
the invariant mass spectra recoiling against $\pi^+\pi^-$.
The following systematic uncertainties of the fit are considered:
$(a)$ \emph{the~fit}: we propagate the statistical uncertainties of the $J/\psi$ signal yield from the
fit to the selection efficiency, and the resulting uncertainties, 0.09\% and
0.08\% for $\epsilon^{\psi(3686)}_{data}$ and $\epsilon^{\psi(3686)}_{MC}$
respectively, are considered to be the uncertainty from the fit itself.
$(b)$ \emph{the~fit~range}: we change the fit range on the $\pi^+\pi^-$ recoiling mass
from [3.07, 3.13]  GeV/$c^2$ to [3.08, 3.12] GeV/$c^2$, and the resulting difference,
0.08\% is taken as a systematic uncertainty. $(c)$ \emph{the~signal~shape}: we perform
an alternative fit by describing the $J/\psi$ signal
with a histogram obtained from the recoil mass spectrum of $\pi^+ \pi^-$ in
$\psi(3686)\rightarrow \pi^+ \pi^- J/\psi, ~J/\psi \rightarrow
\mu^+\mu^-$, and the resulting change, 0.12\%, is considered to be the associated systematic
uncertainty. $(d)$ \emph{the~background~shape}: the uncertainty due to the background shape,
0.02\%, is estimated by replacing the second order Chebychev polynomial with a
first order Chebychev polynomial. By assuming all of the sources of systematic
uncertainty are independent, the fit uncertainty for the 2012 $J/\psi$ sample,
0.19\%, is obtained by adding all of the above effects in quadrature.

The same sources of systematic uncertainty are considered for the $J/\psi$ sample
taken in 2009.  The fit has an uncertainty of 0.02\% for
$\epsilon^{\psi(3686)}_\text{data}$ and 0.07\% for $\epsilon^{\psi(3686)}_\text{MC}$. The
uncertainty from the fit range, signal function and background shape
are 0.02\%, 0.15\% and 0.02\%, respectively. The total uncertainty
from the fit for the 2009 data is 0.17\%.

\subsection{Background uncertainty}
\label{bkgerror}
In the measurement of the number of $J/\psi$ events, the background
events from QED processes, cosmic rays, and beam-gas interactions
are estimated by normalizing the number of events in the continuum
data sample taken at $\sqrt{s} = 3.08$ GeV according to Eq.~(\ref{Nbg}). Therefore the background
uncertainty mainly comes from the normalization method, the statistics of
the sample taken at $\sqrt{s} = 3.08$ GeV, the statistical uncertainty of
the integrated luminosity and the uncertainty due to beam associated backgrounds.

In practice, Eq.~(\ref{Nbg}) is improper for the normalization of the
background of cosmic rays and beam-gas.
The number of cosmic rays is proportional to the time of data taking, while
beam-gas interaction backgrounds are related to the vacuum status and
beam current during data taking in addition to the time of data taking.
Assuming a stable beam and vacuum status, the backgrounds of cosmic rays and
beam-gas interactions are proportional to the integrated luminosity.
Therefore, the difference in the estimated number of backgrounds between that
with and without the energy-dependent factor in Eq.~(\ref{Nbg}) is considered to be the
associated systematic uncertainty.

  In 2012, two data samples with $\sqrt{s} = 3.08$ GeV were taken at the
beginning and end of the $J/\psi$ data taking. To estimate
the uncertainty of the background related with the stability of the beam and
vacuum status, we estimated the background with Eq.~(\ref{Nbg}) for the two
continuum data samples, individually. The maximum change in the nominal
results, 0.05\%, is taken as the associated systematic uncertainty.
In the background estimation for data taken in 2009, only one continuum data
sample was taken. The  corresponding uncertainty is estimated by comparing
the selected background events from the continuum sample to that from the $J/\psi$
data, which is described in detail in ~\cite{njsi2009}.

After considering the above effects, the uncertainties on the number of
$J/\psi$ events related to the background are 0.06\% and 0.13\% for the data
taken in 2012 and 2009, respectively. The uncertainties are determined from the quadratic sum of
the above individual uncertainties, assuming all of them to be independent.

\subsection {Noise mixing}
In the BESIII simulation software, the detector noise and machine
background are included in the MC
simulation by mixing the simulated events with events recorded by a random trigger.
To determine the systematic uncertainty associated with the
noise realization in the MC simulation, the $\psi(3686)$ MC sample is reconstructed
by mixing the noise sample accompanying the $J/\psi$ data taking. The change of
the correction factor for the detection efficiency, 0.09\%, is taken as the systematic
uncertainty due to noise mixing for the number of $J/\psi$ events taken in 2012.

  In the determination of the number of $J/\psi$ events collected in 2009,  106
million of $\psi(3686)$ events taken in 2009 are used to determine the
detection efficiency, and the corresponding uncertainty related to the noise realization is
estimated to be 0.10\% with the same method. However, the noise level was
 not entirely stable during the time of the $\psi(3686)$ data taking. To
check the effect on the detection efficiency related to the different noise levels,
the $\psi(3686)$ data and the MC samples are divided into three sub-samples,
and the detection efficiency and the correction factor are  determined for
the three sub-samples individually. The resulting maximum change in the number of $J/\psi$ events,
0.06\%, is taken as an additional systematic uncertainty associated with the noise realization.
The total systematic uncertainty due to the noise  is estimated to be
0.12\% for the $J/\psi$ events taken in 2009.

\subsection {\boldmath Selection efficiency uncertainty of two soft pions}
According to the MC study, the selection efficiency of soft pions,
$\epsilon_{\pi^+\pi^-}$, recoiling against the $J/\psi$ in
$\psi(3686)\rightarrow \pi^+\pi^-J/\psi$ is found to depend  on the
multiplicity of the $J/\psi$ decays. Differences between
the data and MC samples may lead to a change in the number of $J/\psi$ events.
We compare the multiplicity distribution of $J/\psi$ decays in the $\psi(3686)
\rightarrow \pi^+\pi^- J/\psi$ data sample to that of the $J/\psi$ data at rest
 to obtain the dependence of $\epsilon_{\pi^+\pi^-}$ in the data. The efficiency
determined from the $\psi(3686) \rightarrow \pi^+ \pi^- J/\psi(J/\psi\rightarrow
\text{inclusive})$ MC sample, $\epsilon^{\psi(3686)}_\text{MC}$ in Eq.~(\ref{Fcor}), is reweighted with
the dependence of $\epsilon_{\pi^+\pi^-}$ from the data sample.
The resulting change in the number of $J/\psi$ events, 0.28\% (0.34\%) is taken as the uncertainty
for the data taken in 2012 (2009).

The systematic uncertainties from the different sources studied above are summarized
in Table~\ref{TABSYS}. The total systematic  uncertainty for the number of
$J/\psi$ in 2012 (2009), 0.55\% (0.63\%), is the quadratic sum of the individual uncertainties.
\begin{center}
\tabcaption{\label{TABSYS}Summary of systematic sources and the corresponding
contributions on the number of $J/\psi$ events, where the superscript *
 means the error is common for the data samples taken in 2009 and 2012.}
\footnotesize
\begin{tabular*}{70mm}{l@{\extracolsep{\fill}}cc}
\toprule Sources & 2012 (\%) & 2009(\%)\\ \hline
$^*$MC model uncertainty &0.42 &0.36 \\
Track reconstruction efficiency & 0.03&0.30 \\
Fit to $J/\psi$ peak & 0.19&0.17 \\
Background uncertainty & 0.06&0.13 \\
Noise mixing & 0.09&0.12 \\
$^*$$\epsilon_{\pi^+\pi^-}$ uncertainty &0.28& 0.34 \\\hline
Total & 0.55&0.63 \\
\bottomrule
\end{tabular*}
\end{center}

\section{Summary}
Using inclusive $J/\psi$ events, the number of $J/\psi$
events collected with the BESIII detector in 2012 is determined to be
$ N_{J/\psi2012}= (1086.9\pm6.0)\times10^{6},$
where the uncertainty is systematic only and the statistical uncertainty is negligible.
The number of $J/\psi$ events taken in 2009 is recalculated to be
$ N_{J/\psi2009}= (223.7\pm1.4)\times10^{6},$
which is consistent with the previous measurement~\cite{njsi2009}, but with
improved precision.

In summary, the total number of $J/\psi$ events taken with BESIII detector
is determined to be
$ N_{J/\psi}= (1310.6\pm7.0)\times10^{6}.$
Here, the total uncertainty is determined by adding the common uncertainties directly and
the independent ones in quadrature.

\vspace{6mm}

{\it  The BESIII collaboration thanks the staff of BEPCII and the
IHEP computing center for their hard efforts.}

\end{multicols}

\clearpage

\vspace{6mm}


\begin{thebibliography}{90}

\vspace{3mm}

\bibitem{jsi1}J. J. Aubert et al.  Phys. Rev. Lett., 1974 {\bf 33}:1404
\bibitem{jsi2}J. E. Augustin et al. Phys. Rev. Lett., 1974, {\bf 33}:1406
\bibitem{bes3}Ablikim M et al. Nucl. Instrum. Methods A, 2010, {\bf 614}: 345-399
\bibitem{njsi2009}Ablikim M et al. Chin. Phys. C, 2012, {\bf 36}: 915-925
\bibitem{kkmc} Jadach S, Ward B F L, Was Z, Comput. Phys. Commu. 2000, {\bf
130}:130; Jadach S, Ward B F L, Was Z, Phys. Rev. D, 2001, {\bf 63}:113009.
\bibitem{evtgen} Ping R G, HEP \& NP, 2008, {\bf 32}: 599-602
\bibitem{djl} D. J. Lange, Nucl. Instrum. Methods A, 2001, {\bf 462}:152-155
\bibitem{PDG}   Olive K A et al. (Particle Data Group). Chin. Phys. C, 2014, {\bf 38}: 090001
\bibitem{LUND} Chen J C et al. Phys. Rev. D, 2000, {\bf 62}: 1-8
\bibitem{LUND2} Yang R L, Ping R G, Chen H, Chin. Phys. Lett., 2014, {\bf
31}:061301
\bibitem{geant4} Agostinelli S et al.,  Nucl. Instrum. Methods A, 2003, {\bf
506}:250-303
\bibitem{trig} Berger N et al., Chin. Phys. C, 2010, {\bf 34}:1779-1784
\end{thebibliography}
\end{document}